\documentstyle[psfig,12pt]{article}
\begin{document}

\begin{center}
{\bf A new parallel TreeSPH code}\footnote{To appear in "Science and 
Super Computing at Cineca-Report 1997"} 
\end{center}

\begin{center}
{\bf Cesario Lia$^{1}$, Giovanni Carraro$^{1,2}$, Cesare Chiosi$^{2}$ and 
Marco Voli$^{3}$} 
\end{center}

\noindent
$^{1}$ SISSA/ISAS, via Beirut 2, 34013 Trieste \\
$^{2}$ Dipartimento di Astronomia, Universita' di Padova,
Vicolo dell'Osservatorio 5, 35122 Padova\\
$^{3}$ CINECA, Via Magnanelli 6/3, Casalecchio di Reno, 40033 
Bologna\\
{\bf e-mail}: liac@sissa.it, carraro@pd.astro.it, 
chiosi@pd.astro.it, voli@cineca.it

\begin{center}
{\bf Abstract}
\end{center}
\noindent
In this report we describe a parallel implementation of a
Tree-SPH code realized using the SHMEM libraries in the Cray T3E
supercomputer at CINECA. We show the result of a 3D test to check the
code performances against its scalar version. Finally we compare the load
balancing and scalability of the code with PTreeSPH (Dav\'e et al 1997),
the only other parallel Tree-SPH code present in the literature.

\section{INTRODUCTION}
\noindent
In Numerical Astrophysics there is nowadays a great need to increase 
the dynamical range of N-body simulations. This is mainly the case 
of large Scale Structures and Galaxy Formation simulations, the
motivation being that one would be likely able to simulate volumes
comparable to the dimension of the Universe, but resolving at the same
time the regions in which many important physical processes -like
star formation- occur. The involved dynamical range is of the order of 
$10^{9}$.
The last decade has seen a huge effort to realize good codes
-both eulerian and lagrangian- to study Galaxy Formation.
In particular lagrangian codes, due to their spatial flexibility,
have become a major tool in this kind of applications (Carraro, Lia and
Chiosi 1998).
In the case of lagrangian codes, the dynamical range is improved by using
greater and greater numbers of particles. This fact also has the advantage
to
decrease the errors (which are poissonian) in the model predictions.
Unfortunately scalar (and even vector) particle codes are not suitable
to study with sufficient accuracy Galaxy Formation processes, because
the dynamical range reachable is of the order of some thousands, using
roughly half a million particles. Moreover such big simulations require
enormous computing time.
Therefore the  use of massively supercomputers is the possible way out
for these resolution problems.
The first example of the parallelization  of a particle code is PTreeSPH
(Dav\'e, Dubinski and Hernquist 1998). This code allows to run simulations
with half a million particles in the reasonable time of 6000 node-hours.
PTreeSPH is a C code which uses MPI communication package in a Cray T3D
machine.
In this report we describe a new implementation of a TreeSPH code realized
using SHMEN libraries for the 128 nodes Cray T3E supercomputer hosted by 
CINECA. We are going to use this code to study individual galaxy
formation and evolution. This approach allows us to decrease by
several order the necessary dynamical range and to perform useful
simulations by using some hundreds thousand  particles.

\begin{figure*}
\centerline{\psfig{file=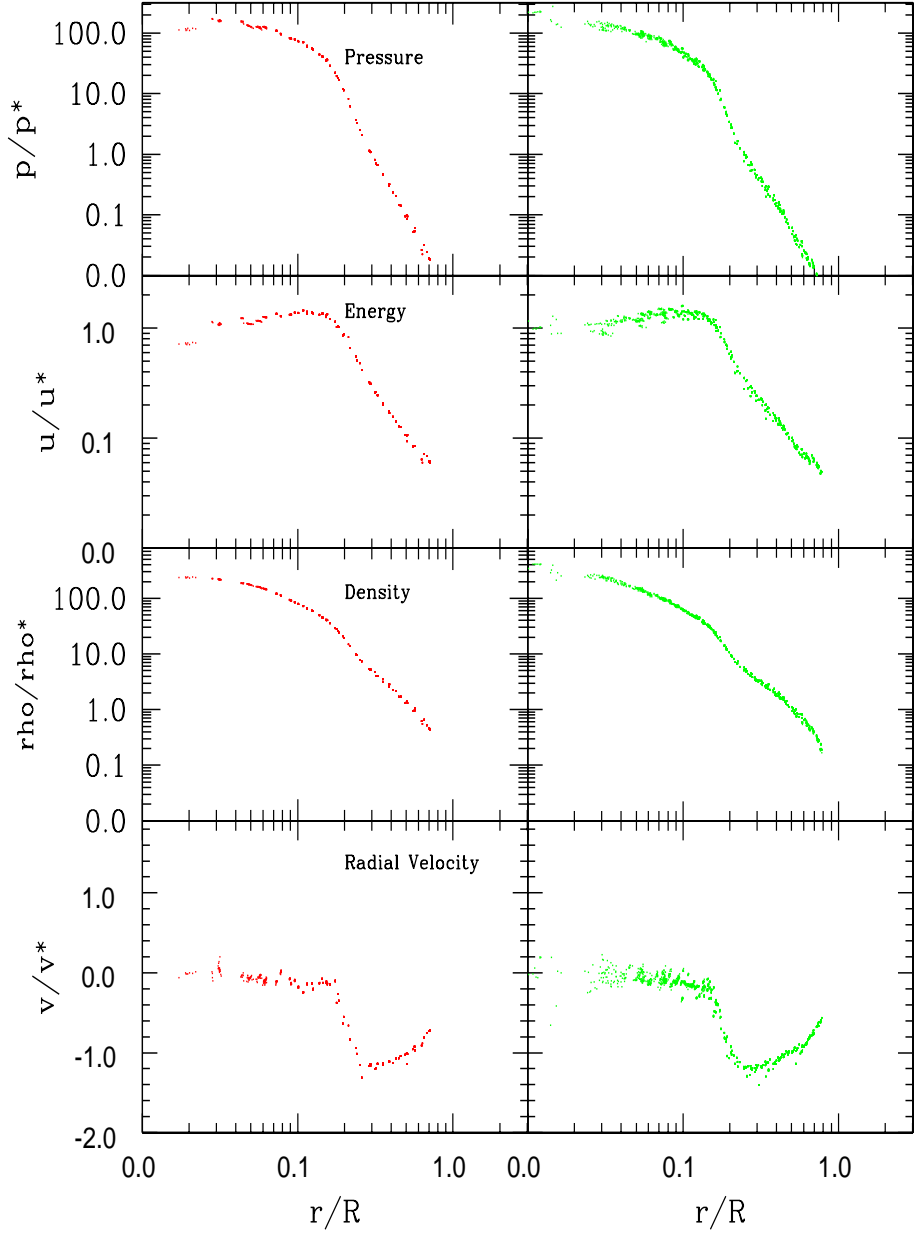,height=17cm,width=14cm}}
\caption[]{Adiabatic collapse of an isothermal gas sphere. Snapshots refer
to t = 0.88 when the shock is most prominent. Left panels refer to the 
scalar version,right to the parallel one.} 
\end{figure*}

\section{THE CODE}
\noindent
Our TreeSPH is a F90 code, which is  described in details in Carraro, Lia
and Chiosi (1998). It is a gravity treecode  combined with an SPH
hydrodynamics. It uses variable smoothing lengths and individual time
 steps.
The scalar version of the code contains a very good
implementation of the baryons physics, i.e. cooling, star formation, 
feedback and chemical evolution (Lia, Carraro and Chiosi 1998;
Carraro, Lia and Buonomo 1998).

To keep processors balanced, the parallel code utilizes a domain
decomposition algorithm based on a
particle workload criterion. At any time step we compute the particle
workload on the basis of the particle timestep, which gives and idea of
the amount of floating point operations (gravity plus hydro) of the
particle itself. The procedure consists in  deriving the center of mass of
the workloads
distribution and iteratively subdivide the volume distributing particles
to the processors. The subdivisions are not necessarily performed along
planes parallel to the reference frame axis, the only real goal being to
render sub-volumes as more spherical as possible.

Communications are handled using Shared Memory paradigm with the
SHMem-library
routines. This choice is advantageous in a number or way. First, these
routines
are provided by Cray Research itself to have the best use of the hardware 
remote memory access available on the Cray T3E (which is the CINECA's machine
we used for our work); thus we have in turn the best performance for
communication.
Second,  having no more need of syncronization for communication (an
explicit
receive for each sent message) our algorithm greatly gains in semplicity
and 
the synchronization performance cost is eliminated too.
The only drawback might be the portability, but the new version 2.0 of the
standard MPI library already includes the On-Side Communication paradigm
which implements a remote memory access much similar to the Shared Memory;
this will allow us to easily port our program to the MPI 2 coding style,
which should
naturally grow in popularity thanks to the popularity already gained by
its predecessor (MPI 1).

Communications are controlled using a data structure made of two trees.
Any CPU builds at any timestep its own tree up, then it descends once all
other trees to gather all the necessary information to compute SPH
estimates of the physical quantities. We call this structure ghost
tree. This has some similarities with the
local essential tree. The descent allows simultaneously to compute gravity
and to perform neighbors searching. Care is also payed to the memory
necessary to maintain remote informations. In addition, if the domain
decomposition is made efficiently, it is not necessary to descend entirely
a remote tree, this descend being decided in obey to an opening criterion
similar to the local tree opening criterion.

\begin{figure}
\centerline{\psfig{file=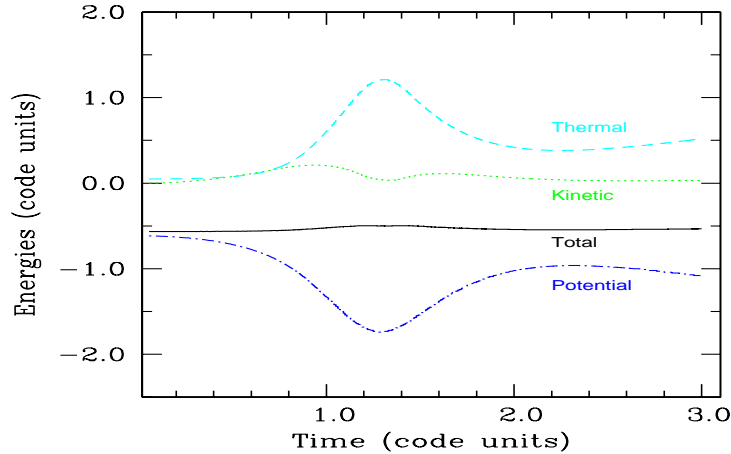,height=8cm,width=12cm}}
\caption[]{Energy conservation in a run with 16 processors.} 
\end{figure}

\section{A 3-DIMENSIONAL TEST}
\noindent
In order to analyze the behavior and the performances of the code,
we have run the adiabatic collapse test, which is an ideal test to check
the response of the code to situations in which a large dynamic range
in space and time is present. The comparison between the results of the
scalar and parallel version is presented in Fig. 1, where left panels
refer to the scalar version(red), right panels to the parallel
one(green).
The collapsing gas
produces a shock most prominently at $t = 0.88$ (time is in code units,
where total mass M, total radius R and gravitational constant G are set
equal to 1), and for this time we show
the comparison of density, pressure, internal energy and radial velocities
profiles. This tests has been realized using 4000 particles distributed
in 16 processors. The scalar test has been realized using 2176 particles.
The results are quite good, except for the inner region
of the sphere, which by the way is under the resolution, which is governed
by the gravity softening parameter, kept equal to 0.1 for consistency
with Dav\'e, Dubinski and Hernquist 1998.
In Fig.2 we show energy conservation, which in this case is under $8\%$.

\begin{figure*}
\centerline{\psfig{file=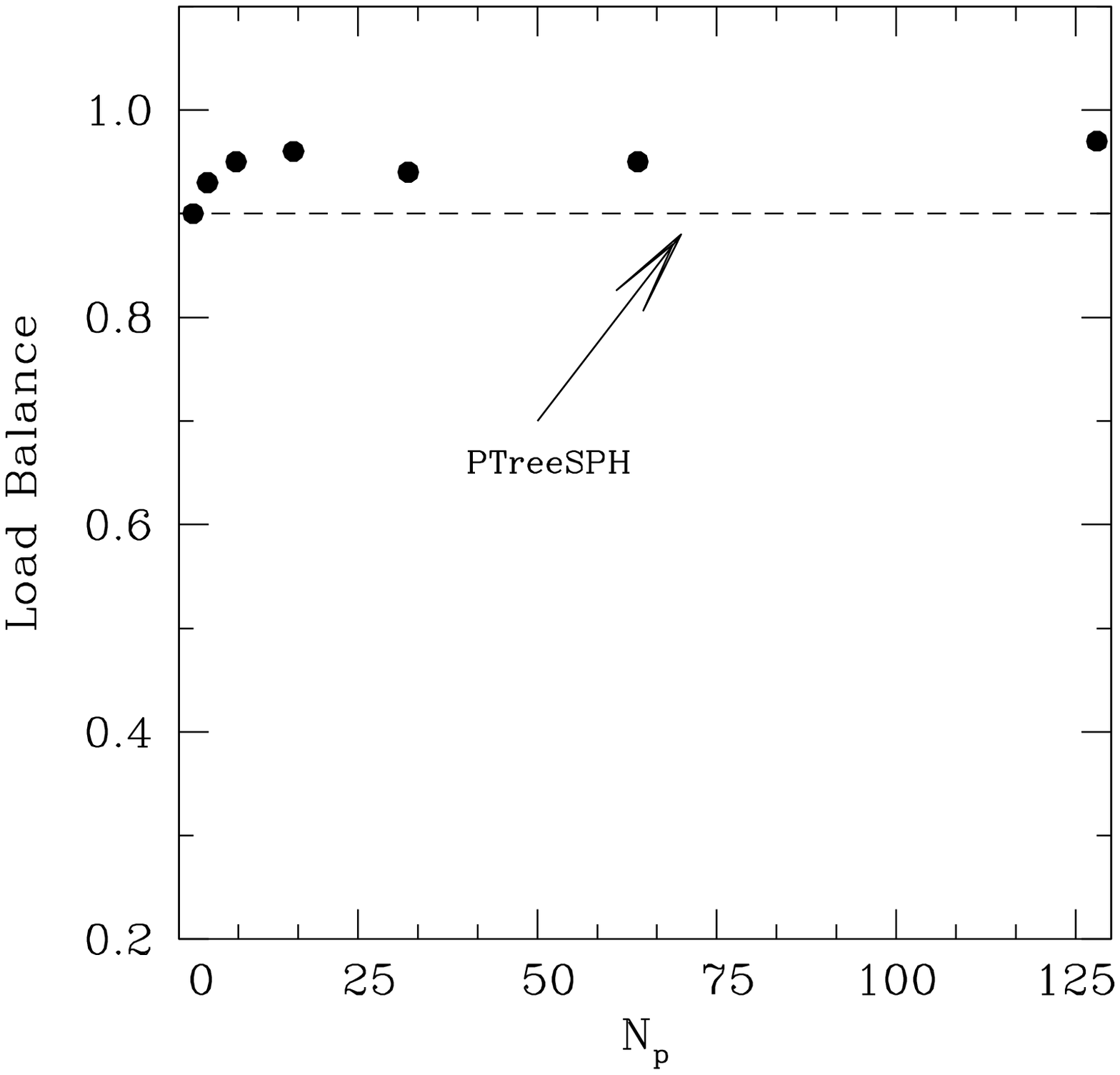,height=14cm,width=14cm}}
\caption[]{Global Load Balance as given by equation 1. Dashed line
represents PTreeSPH mean load balance.} 
\end{figure*}

\begin{figure*}
\centerline{\psfig{file=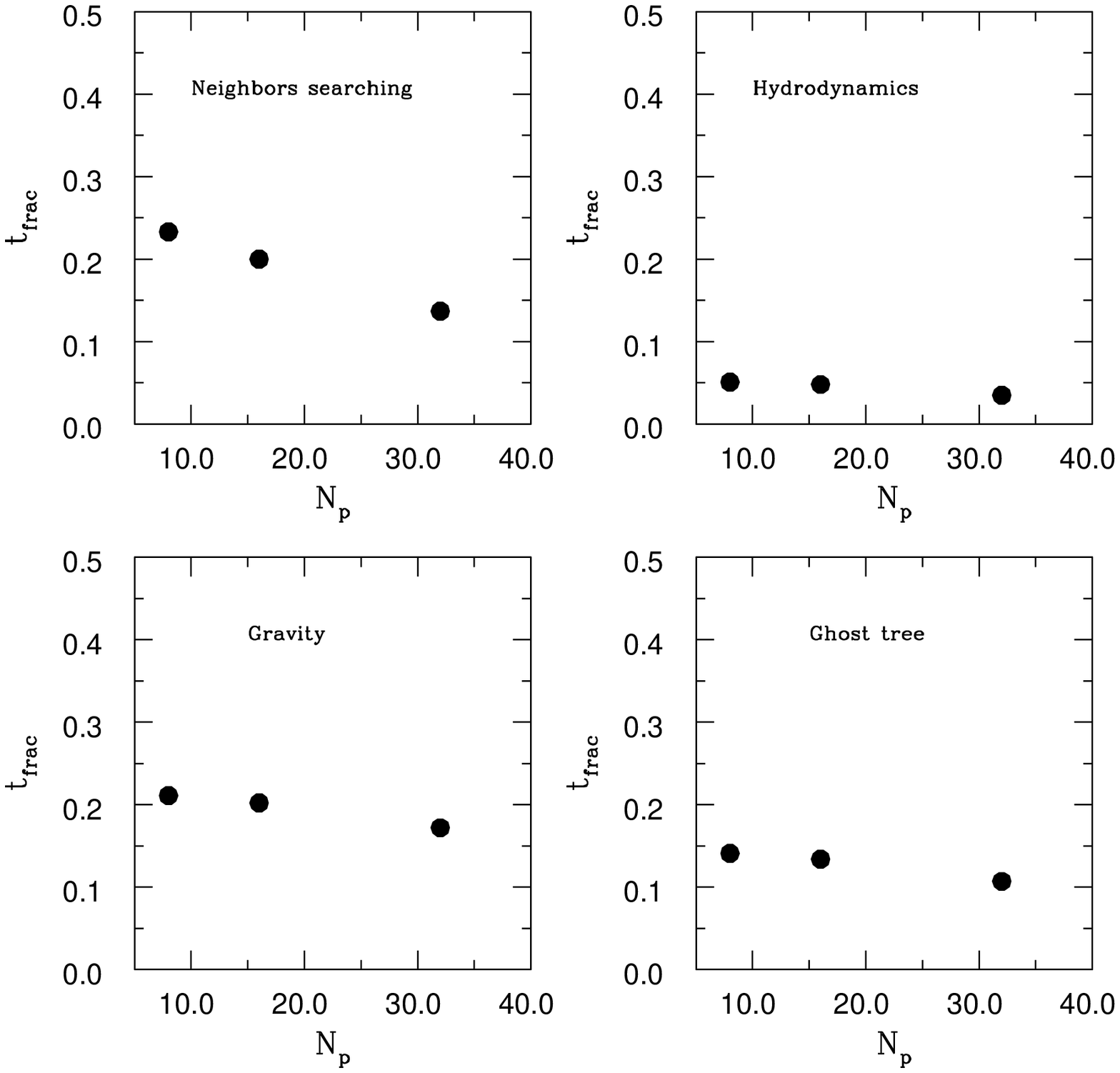,height=16cm,width=14cm}}
\caption{Scalability of four subroutines using 8, 16, and 32
processors.} 
\end{figure*}

\section{PERFORMANCE ANALYSIS}
\noindent
In this section we present the performance of the code in terms
of load balancing and scalability. To this aim we performed the adiabatic
collapse test at increasing number of processors, and computed
the load balance according to Dave', Dubinski and Hernquist (1998)
equation (12):

\begin{equation}
L = \frac{1}{N_{p}} \sum_{i=1}^{N_{p}} 1 - \frac{t_{max} - t_{i}}{t_{max}}
\end{equation}

The result is plotted in Fig. 3, where dashed line
shows the mean load balance for the PTreeSPH code.
Our global load balance is always greater than $90\%$, which means
that the time a processor remains idle is always lower than $10\%$. This is
mainly due to the asynchronous communications scheme used.
Scalability means that any routine in the code should speed up linearly
at increasing number of processors. Here we consider four routines:
gravity, neighbors searching, hydrodynamics and ghost tree construction.
The scalability is measured in terms of the time fraction spent by any
subroutine. Fig. 4 shows that gravity, neighbors searching and ghost
tree
construction subroutines scale linearly with the number of processors.
The computation of hydrodynamics does not decrease too much, this being
less than a problem, due to the small time spent by this subroutine.

\section{FUTURE PERSPECTIVES}
\noindent
At present our parallel code contains also an implementation of cooling
phenomena. We are going to implement also star formation, feedback and
chemical evolution (Lia, Carraro, Chiosi and Voli 1998). 
Some additional work is in progress to render the code more efficient
implementing variable softening parameters and anisotropic smoothing
lengths. By using half a million particles we hope to resolve regions of
$10^{4}$ solar masses in a $10^{11}$ solar masses galaxy, or to easily 
resolve the stellar distribution in a $10^{6}$ solar masses globular 
cluster.\\

\noindent
{\bf Aknowledgements}
\noindent
This work has been realized thanks to a T3E grant given by CINECA~ to Cesare 
Chiosi.  C. Lia and G. Carraro deeply thank Proff. Cesare Chiosi and Luigi 
Danese for their constant and enthusiastic support.

\section{BIBLIOGRAPHY}

Carraro G., Lia C., Chiosi C., 1998 ({\tt astro-ph/9712307})\\
Carraro G., Lia C., Buonomo F., 1998, proceedings of "DM-ITALIA 1997", 
Trieste December 9-11 1997,in press\\
Dave' R., Dubinski J., Hernquist L., 1997 ({\tt astro-ph 9701113})\\
Lia C., Carraro G., Chiosi C., 1998, MNRAS, submitted\\
Lia C., Carraro G., Chiosi C., Voli M., 1998, MNRAS in preparation

\end{document}